\documentclass{ws-procs975x65}

\begin{document}

\title{\uppercase{Can one observe quantum-gravitational effects in the Cosmic Microwave Background?}}

\author{MANUEL KR\"AMER}

\address{Institut f\"ur Theoretische Physik, Universit\"at zu K\"oln, \\
Z\"ulpicher Stra{\ss}e 77, 50937 K\"oln, Germany\\
E-mail: mk@thp.uni-koeln.de}

\begin{abstract}
In order to find the correct theory of quantum gravity, one has to look for observational effects in any candidate theory. Here, we focus on canonical quantum gravity and calculate the quantum-gravitational contributions to the anisotropy spectrum of the cosmic microwave background that arise from a semiclassical approximation to the Wheeler--DeWitt equation. While the resulting modification of the power spectrum at large scales is too weak to be observable, we find an upper bound on the energy scale of inflation.
\end{abstract}


\bodymatter

\section{Introduction}

Up to today, various approaches of a theory of quantum gravity have been developed, but none of them has lead to a testable prediction yet. The main reason for this failure is the general smallness of effects in which both gravity and quantum theory play a role. It is expected that quantum gravity will only become sizable in situations involving energies in the region of the Planck scale, i.e.~of approximately $10^{19}\,\text{GeV}$, which in principle restricts the search for quantum-gravitational effects to black hole physics and cosmology of the very early universe.

Here, we want to restrict ourselves to looking for potentially observable effects of quantum gravity in the  cosmic microwave background (CMB) radiation. As a framework we use \emph{Quantum Geometrodynamics}, a direct canonical quantization of general relativity, and we calculate the dominant quantum-gravitational contribution to the CMB anisotropy spectrum by performing a semiclassical approximation to the Wheeler--DeWitt equation of a suitable quantum-cosmological model. \cite{KK12, grf}

\section{The quantum-cosmological model}

We consider an inflationary universe with perturbations of a scalar field $\phi$. The background universe is assumed to be a flat Friedmann--Lema\^itre universe with scale factor $a\equiv\exp(\alpha)$, where $\phi$ plays the role of the inflaton with the potential $\mathcal{V}(\phi) = \frac{1}{2}\,m^2\phi^2 \approx \mathrm{const.}$ We also demand the slow-roll approximation $\dot{\phi}^2 \ll \vert\mathcal{V}(\phi)\vert$ to hold. The perturbations of the inflaton field are decomposed into Fourier modes $\delta\phi(\mathbf{x},t) = \sum_k f_k(t)\,\mathrm{e}^{\mathrm{i}\mathbf{k}\cdot\mathbf{x}}$ with the wave number $k$ defined to be dimensionless.

With the convention $\hbar=c=1$, a redefinition of the Planck mass as $m_{\rm P}:=\sqrt{3\pi/2G}\approx 2.65 \times 10^{19}\,\mathrm{GeV}$ and the quasi-static Hubble parameter denoted as $H$, the Wheeler--DeWitt equation for each of the modes $f_k$ takes the form \cite{Hall85}
\begin{equation} \label{wdw}
\bigl[\mathcal{H}_0+\mathcal{H}_{k}\bigr]\Psi_k\!\left(\alpha,f_k\right)=0\,,
\end{equation}
where $\mathcal{H}_0$ and $\mathcal{H}_k$ are the Hamiltonians  of the background and the perturbation modes, respectively:
\[
{\mathcal H}_0 = \frac{\mathrm{e}^{-3\alpha}}{2}\left[\frac{1}{m_{\rm P}^2}\frac{\partial^2}{\partial\alpha^2}+\mathrm{e}^{6\alpha}m_{\rm P}^2H^2\right], \;\; \mathcal{H}_k =
\frac{\mathrm{e}^{-3\alpha}}{2}\left[-\,\frac{\partial^2}{\partial
    f_k^2} + \Bigl(k^2\mathrm{e}^{4\alpha} +
  m^2\mathrm{e}^{6\alpha}\Bigr)f_k^2\right] .
\]

\section{Derivation of the power spectrum}

We perform a semiclassical approximation \cite{KS91} to equation \eqref{wdw}  by making the ansatz $\Psi_k(\alpha,f_k) = \mathrm{e}^{\mathrm{i}\,S(\alpha,f_k)}$ and expanding $S$ in terms of powers of $m_{\rm P}^2$:
\[
S(\alpha,f_k) = m_{\rm P}^2\,S_0 + m_{\rm P}^0\,S_1 + m_{\rm P}^{-2}\,S_2 + \ldots
\]
Inserting this ansatz into equation (\ref{wdw}) and comparing terms of equal power of $m_{\rm P}$, we recover the Hamilton--Jacobi equation for the classical minisuperspace background at order $\mathcal{O}(m_{\rm P}^{2})$. At the order $\mathcal{O}(m_{\rm P}^{0})$, we define $\psi^{(0)}_{k}(\alpha,f_{k})\equiv
\gamma(\alpha)\,\mathrm{e}^{\mathrm{i}\,S_{1}(\alpha,f_{k})}$ with $\gamma(\alpha)$ imposed to be equal to the standard WKB prefactor. Now we can introduce a time parameter $t$ that arises from the minisuperspace background:
\[
\frac{\partial}{\partial t} :=
-\,\mathrm{e}^{-3\alpha}\frac{\partial S_0}{\partial
    \alpha}\,\frac{\partial}{\partial \alpha}\,.
\]
We then find that each $\psi^{(0)}_{k}(\alpha,f_{k})$ obeys a Schr\"odinger equation with respect to $t$:
\begin{equation} \label{schrgl}
\mathrm{i}\,\frac{\partial}{\partial t}\,\psi^{(0)}_{k}=\mathcal{H}_{k} \psi^{(0)}_{k}\,.
\end{equation}
In order to obtain the power spectrum of the scalar field perturbations, we solve equation \eqref{schrgl} by using the Gaussian ansatz 
$\psi^{(0)}_{k}(t,f_k) =
\mathcal{N}_k^{(0)}(t)\,\mathrm{e}^{-\frac{1}{2}\,\Omega_k^{(0)}(t)\,f_k^2}$.
The density contrast $\delta_k(t)$ in the slow-roll regime at the time $t_\mathrm{enter}$ when the respective mode $k$ reenters the Hubble radius is defined as
\[
\delta_k(t_\mathrm{enter}) \propto \left|\frac{\mathrm{d}}{\mathrm{d}t}\,\Re\mathfrak{e}\!\left[\Omega_k^{(0)}(t)\right]^{-1/2}\right|_{t_{\mathrm{exit}}}\,.
\]
By inserting the solution for $\Omega_k^{(0)}(t)$, we find that $\delta_k(t_\mathrm{enter}) \propto k^{-3/2}$, which leads to an approximately scale-invariant power spectrum
\[
\mathcal{P}^{(0)}(k) \propto k^{3}\left|\delta_{k}(t_{\rm enter})\right|^{2} \propto H^4\,\big|\dot{\phi}(t)\big|^{-2}_{t_{\mathrm{exit}}} \approx \text{const.}
\] 

\section{Quantum-gravitational corrections}

If we take the semiclassical approximation one step further, to the order $\mathcal{O}(m_{\rm P}^{-2})$, we can define the wave functions $\psi^{(1)}_{k}(\alpha,f_k)$ that
obey a quantum-gravitationally corrected Schr\"odinger equation of the form
\begin{equation} \label{corr_Schr_eq}
\mathrm{i}\,\frac{\partial}{\partial t}\,\psi^{(1)}_{k} =
\mathcal{H}_{k}\psi^{(1)}_{k} - \frac{\mathrm{e}^{3\alpha}}{2m_{\rm P}^2\psi^{(0)}_k}\biggl[\frac{\bigl(
\mathcal{H}_{k}\bigr)^2}{V}  
\psi^{(0)}_{k} + \mathrm{i}\,\frac{\partial}{\partial t}
\left(\frac{\mathcal{H}_{k}}{V}\right)
\psi^{(0)}_{k}\biggr]\psi^{(1)}_k\,,
\end{equation}
where $V := \mathrm{e}^{6\alpha}\,H^2$. Since the first term in the brackets gives the dominant contribution, we will neglect the second term in the following.

From equation (\ref{corr_Schr_eq}), we can calculate the quantum-gravitational correction to the power spectrum derived above by using the following modified Gaussian ansatz:
\[
\ \ \psi^{(1)}_{k}(t,f_k) = \left(\mathcal{N}_k^{(0)}(t) +
  \frac{1}{m_{\rm P}^2}\,\mathcal{N}_k^{(1)}(t)\right) \exp\!\left[-\,\frac{1}{2}\left(\Omega_k^{(0)}(t)+\frac{1}{m_{\rm
        P}^2}\,\Omega_k^{(1)}(t)\right)f_k^2\right]  .
\]
With the boundary condition $\Omega_k^{(1)}(t) \rightarrow 0$ as $t \rightarrow \infty$, we find that we can incorporate the quantum-gravitational modification into a correction term $C(k)$ relating the uncorrected power spectrum ${\cal P}^{(0)}$ to the corrected one ${\cal P}^{(1)}$ via ${\cal P}^{(1)}(k)={\cal P}^{(0)}(k)\,C(k)$:
\[
C(k)=1+ \frac{\delta^{\pm}}{k^3}\,\frac{H^2}{m_{\mathrm{P}}^2} + \frac{1}{k^6}\,{\mathcal O}\!\left(\frac{H^4}{m_{\mathrm{P}}^4}\right) . 
\]
While at first we determined the prefactor $\delta^\pm \in \mathbb{R}$ to take the value $\delta^- = -\,247.68$,\cite{KK12} a recent investigation has found that a subtle change in the implementation of the boundary condition can lead to the alternative value $\delta^+ = 179.09$.\cite{BEKKP13} Apart from the sign change, the order of the correction is the same in both cases, however.

One thus sees that the quantum-gravitational correction explicitly breaks the scale invariance of the uncorrected power spectrum and leads to a modification of the power at large scales (small $k$). However, the quantum-gravitational effect would only become significant if the inflationary Hubble parameter $H$ were close to the Planck scale. But since one can derive an upper bound on $H$ from the observational bound of the scalar-to-tensor ratio, $H \lesssim 4\times 10^{-6}\,m_{\mathrm{P}} \approx 10^{14}\,\mathrm{GeV}$, we have to conclude that even in this limiting case the quantum-gravitational effect is extremely small. Furthermore, measurement accuracy is fundamentally limited at large scales by cosmic variance. This implies that one will not be able to observe this effect even with more precise measurements of the CMB anisotropies by e.g.~PLANCK.\cite{Cal12}

Nevertheless, we can use our analysis to independently derive an upper bound on the Hubble parameter. Since it was observed that the power spectrum deviates by less than $5\,\%$ from a scale-invariant spectrum \cite{Koma11}, we assume that $C(k)$ has to be within the limits $0.95 \lesssim C(k) \lesssim 1.05$ for $k \sim 1$. From this condition we deduce the weaker bound $H \lesssim 1.5\times 10^{-2}\,m_{\mathrm{P}} \approx 4\times10^{17}\,\mathrm{GeV}$. \\

\noindent {\bf Acknowledgements.} The author thanks Claus Kiefer for useful remarks and gratefully acknowledges support from the Bonn--Cologne Graduate School of Physics and Astronomy.


\begin{thebibliography}{9}
\bibitem{KK12} C. Kiefer and M. Kr\"amer, {\em Phys. Rev. Lett.} {\bf 108}, 021301 (2012).
  
\bibitem{grf} C. Kiefer and M. Kr\"amer, {\em Int. J. Mod. Phys. D} {\bf 21}, 1241001 (2012).

\bibitem{Hall85} J.~J. Halliwell and S.~W. Hawking, {\em Phys. Rev. D} {\bf 31}, 1777 (1985).

\bibitem{KS91} C. Kiefer and T.~P. Singh, {\em Phys. Rev. D} {\bf 44}, 1067 (1991).

\bibitem{BEKKP13} D. Bini, G. Esposito, C. Kiefer, M. Kr\"amer, and F. Pessina, arXiv:1303.0531.

\bibitem{Cal12} G. Calcagni, arXiv:1209.0473.

\bibitem{Koma11} E. Komatsu {\em et al.}, {\em Astrophys. J. Suppl. Ser.} {\bf 192}, 18 (2011); G. Hinshaw {\em et al.}, arXiv:1212.5226.

\end{thebibliography}
\end{document}